\documentclass[aps,pre,superscriptaddress,twocolumn,amsmath,amssymb,showpacs]{revtex4}
\usepackage{graphicx}
\usepackage{dcolumn}
\usepackage{bm}

\graphicspath{{./fig/}}  

\begin{document}


\title{ Fractional Fokker-Planck subdiffusion in alternating fields }


\author{E.~Heinsalu}
  \affiliation{National Institute of Chemical Physics and Biophysics, 
  R\"avala 10, 
  Tallinn 15042, Estonia}
  \affiliation{IFISC, 
  Instituto de F\'isica Interdisciplinar y Sistemas Complejos (CSIC-UIB), 
  E-07122 Palma de Mallorca, Spain}

\author{M.~Patriarca}
  \affiliation{National Institute of Chemical Physics and Biophysics, 
  R\"avala 10, 
  Tallinn 15042, Estonia}
  \affiliation{IFISC, 
  Instituto de F\'isica Interdisciplinar y Sistemas Complejos (CSIC-UIB), 
  E-07122 Palma de Mallorca, Spain}

\author{I.~Goychuk}
  \affiliation{Institut f\"ur Physik,
  Universit\"at Augsburg,
  Universit\"atsstr. 1,
  D-86135 Augsburg, Germany}

\author{P.~H\"anggi}
  \affiliation{Institut f\"ur Physik,
  Universit\"at Augsburg,
  Universit\"atsstr. 1,
  D-86135 Augsburg, Germany}

\date{\today}

\begin{abstract}
The fractional Fokker-Planck equation for subdiffusion in time-dependent
force fields is derived from the underlying continuous time random walk.
Its limitations are discussed and it is then applied to the study of 
subdiffusion under the influence of a time-periodic rectangular force.
As a main result, we show that such a force does not affect the universal scaling 
relation between the anomalous current and diffusion when applied to the biased dynamics: 
in the long time limit subdiffusion current and anomalous diffusion are immune to the driving. 
This is in sharp contrast with the unbiased case when the subdiffusion coefficient can be 
strongly enhanced, i.e. a zero-frequency response to a periodic driving is present.
\end{abstract}

\pacs{05.40.-a, 05.40.Fb, 05.60.-k, 02.50.Ey}


\maketitle


\section{Introduction}


The theoretical study of anomalously slow relaxation processes in time-dependent force fields 
constitutes a challenge of current research interest which is not free of ambiguity. 
It is known that there is no unique physical mechanism responsible for the 
occurrence of subdiffusion in condensed media \cite{Bouchaud1990}. 
One possible mechanism, which will be addressed in this work, corresponds to 
disordered glassy-like media consisting of trapping domains where the traveling particle can dwell
for a random time with divergent mean value \cite{Scher, ScherMontroll, shlesinger1974}. 
The successive residence times in traps are assumed to be mutually uncorrelated. 
Diffusion is nevertheless a non-Markovian (semi-Markovian) process exhibiting long
(quasi-infinite) time correlations in the particle positions with a weak ergodicity breaking \cite{BelBarkai}. 
Mathematically this physical picture can be described by a continuous time random walk (CTRW) 
model \cite{ScherMontroll, shlesinger1974} which in the continuous space limit leads
to the fractional Fokker-Planck equation (FFPE) \cite{metzler2000R, letter, heinsalu2006b}. 
This latter formulation is incomplete, as no non-Markovian master equation can define 
the underlying non-Markovian  stochastic process \cite{HT}. 
However, the FFPE is very useful and has a tightly associated, complete description with (ordinary)  
Langevin equation in subordinated, random operational time \cite{Fogedby1994, Stanislavsky, Magdziartz}.

The generalization of the CTRW and FFPE to time-dependent
forces is a highly non-trivial matter since the force changes in
the real and not in the operational time \cite{heinsalu2007b, goychuk07}. 
Also, how a field varying in time affects the distribution of the residence 
times in the traps is not clear without
specifying a concrete mechanism or some plausible model,
especially when the mean residence time does not exist \cite{PRE04}. 
The FFPE describing the dynamics in time-dependent force fields $F(x,t)$  
becomes ambiguous with a frequently (ab)used, \textit{ad hoc} version \cite{sokolov2001, Sample} 
which lacks clear theoretical grounds \cite{heinsalu2007b}. 
The correct version of the FFPE for time-dependent fields was first given in 
Refs.~\cite{SokolovKlafter06, heinsalu2007b}: differently from the FFPE for a 
time-independent force, in the case of a time-dependent field the fractional derivative 
does not stand in front of the Fokker-Planck operator but after it. 
As we explain with this work in more detail, such a FFPE can be justified beyond 
the linear response approximation within a CTRW approach only for a special 
class of dichotomously fluctuating fields.
The derivation of the FFPE for subdiffusion in such time-dependent fields 
is presented in Sec.~\ref{derivation}. 

In Sec.~\ref{results} we apply the derived FFPE to study the influence of 
time-periodic rectangular fields on subdiffusive motion. 
Analytical solutions of the FFPE are confirmed by stochastic 
Monte Carlo simulations of the underlying CTRW.  
In particular, we show with this work that the universal scaling relation between 
the biased anomalous diffusion and sub-current 
\cite{ScherMontroll, shlesinger1974, letter, heinsalu2006b} is not affected by the periodic driving.  
Neither current nor diffusion are influenced asymptotically by the time-periodic field. 
This is in spite of the fact that unbiased subdiffusion of the studied kind can be  strongly
enhanced in the time-periodic field \cite{SokolovKlafter06, heinsalu2007b}.


\section{Derivation of the FFPE for time-dependent fields from the underlying CTRW} \label{derivation}


Since the FFPE does not define the underlying stochastic non-Markovian process, 
its generalization to include the influence of a time-dependent field should 
start from the underlying CTRW \cite{metzler2000R, letter}.
Following the general picture of the CTRW we introduce a
one-dimensional lattice $\{ x_i = i \Delta x\}$ with a lattice period
$\Delta x $ and $i = 0, \pm 1, \pm 2, \ldots$
Let us first assume that there is no time-dependent field.  
After a random trapping time $\tau$ a particle at site $i$ hops with  probability
$q _i ^\pm $ to one of the nearest neighbor sites $i \pm 1$; $q_i^+ + q_i^-=1$.  
The random time $\tau$ is extracted from a site-dependent 
residence time distribution (RTD) $\psi_i (\tau)$. 
The corresponding generalized master equation for
 populations $P_i(t)$ reads \cite{Kenkre1973,Hughes,Weiss}
\begin{eqnarray} \label{GME}
\dot P_i(t) &=& \int_0^t \{K_{i-1}^{+}(t-t') P_{i-1}(t') +
K_{i+1}^{-}(t-t') P_{i+1}(t') \nonumber \\
&-& [K_i^{+}(t-t') + K_i^{-}(t-t')] P_i(t')\} \, \mathrm{d} t' \, .
\end{eqnarray} 
The Laplace-transform of the kernel $K_i^{\pm}(t)$ is related to the Laplace-transform 
of the RTD $\psi_i (\tau)$ via $\tilde K_i^{\pm}(s) = q_i^{\pm} s\tilde \psi_i(s) / [1 - \tilde \psi_i(s)]$. 
In the presence of a time-dependent field the kernels become generally functions 
of both instants of time and not only of their difference, i.e. $K_i^{\pm}(t-t')\to K_i^{\pm}(t,t')$. 
One can relate $K_i^{\pm}(t,t')$ with the corresponding time-inhomogeneous RTDs
 $\psi_i^{\pm}(t+\tau,t)\equiv \psi_i^{\pm}(\tau|t)$, which are conditioned on the entrance time $t$. 
However, one always needs a concrete and physically meaningful model to proceed further \cite{PRE04}. 
A simple example is a Markovian process with time-dependent rates $g_i^{\pm}(t)$, where 
$\psi_i^{\pm}(\tau|t) = g_i^{\pm}(t+\tau) \exp\left \{ -\int_{t}^{t+\tau}[g_i^{+}(t') + g_i^{-}(t')]dt'\right \}$ 
and $K_i^{\pm}(t,t') = 2g^{\pm}_i(t) \, \delta(t-t')$. 
This yields in Eq.~(\ref{GME}) the standard master equation for a 
time-inhomogeneous Markovian process. 
How the non-exponential RTDs will be modified for a time-inhomogeneous 
process is generally not clear \cite{PRE04}.
In the present case, one can assume that the trapping occurs due to the 
existence of  direction(s) orthogonal to the $x$-coordinate. 
According to the modeling in Refs.~\cite{SokolovKlafter06, Magdziartz}, 
an external field directed along $x$ 
would not affect the motion in the orthogonal direction(s).
However, it is not correct to think that the RTD in the trap will not be 
influenced by the field acting in the direction of $x$, as it will change the rates 
(let us assume this simplest, tractable model) 
for moving left or right when escaping from the trap. 
Therefore, the RTD will generally be affected, see e.g. Ref.~\cite{PRE05}.
Obviously, the only situation when the RTD in the trap will not be changed is 
when the sum of the rates to escape from the trap, either left or right, is constant. 
In that case, only the probabilities $q_i^{\pm}(t)$ acquire additional 
time-dependence and not the RTDs $\psi_i(\tau)$. 
This corresponds to the special class of dichotomously fluctuating force fields 
$F (x_i, t) = F (x_i) \xi (t)$, where $\xi (t) = \pm 1$. 
Beyond this class, at most the linear response approximation can work \cite{SokolovKlafter06}. 
Therefore, we restrict our treatment to the above class of fluctuating potentials.
In this case, we can write $K_i^{\pm}(t,t') = q^{\pm}_i(t)K_i(t-t')$,
where $\tilde K_i(s) = s\tilde \psi_i(s) / [1 - \tilde \psi_i(s)]$.
Furthermore, we use the Mittag-Leffler distribution for the residence times \cite{metzler2000R},
\begin{equation} \label{ML}
\psi_i(\tau) = -\frac{\mathrm{d}}{\mathrm{d} \tau} E_{\alpha}[
-(\nu_i \tau)^{\alpha}] \, .
\end{equation}
Here $E_{\alpha}(z) = \sum_{n = 0}^{\infty} z^n / \Gamma(n \alpha + 1)$ denotes 
the Mittag-Leffler function, $\alpha \in (0,1)$ is the index of subdiffusion, and 
$\nu_i = [g_i^+(t) + g_i^-(t)]^{1/\alpha}$ the time scaling parameter; 
$g_{i}^\pm (t) = q^{\pm}_i(t) \nu_i^{\alpha}$.
Then $\tilde K_i(s) = \nu_i^{\alpha} s^{1-\alpha}$ and we get
\begin{eqnarray} \label{FME}
\dot P_i(t) &=& g_{i-1}^+(t)\sideset{_0}{_t} {\mathop{\hat
D}^{1-\alpha}} P_{i-1}(t) + g_{i+1}^-(t) \sideset{_0}{_t} {\mathop{\hat
D}^{1-\alpha}} P_{i + 1}(t)  \nonumber \\
&-& [g_i^+(t) + g_i^-(t)]\sideset{_0}{_t} {\mathop{\hat
D}^{1-\alpha}} P_i(t) \, ,
\end{eqnarray}
where the symbol $\sideset{_0}{_t}{\mathop{\hat D}^{1-\alpha}}$ stands for the 
integro-differential operator of the Riemann-Liouville fractional derivative 
acting on a generic function of time $\chi (t)$ as 
\begin{equation} \label{RL}
\sideset{_0}{_t}{\mathop{\hat D}^{1 - \alpha}} \chi (t) 
= \frac{1}{\Gamma (\alpha)} \frac{\partial}{\partial t} \int_{0}^{t} \mathrm{d} t' \, \frac{\chi (t')}{(t - t')^{1 - \alpha}} \, ;
\end{equation}
$\Gamma (\alpha)$ is the gamma-function.
In a time-dependent potential $U(x,t)$, one can set
\begin{eqnarray} \label{f-rate}
g_i^\pm(t) &=& (\kappa_{\alpha} / \Delta x ^2)
\exp \{ -\beta[U_{i \pm 1/2}(t) - U_i(t)] \} \nonumber \\
&\approx & (\kappa_{\alpha} / \Delta x ^2)
\exp [\pm \beta F(x_i,t)\Delta x/2] \, ,
\end{eqnarray}
so that the Boltzmann relation $g_{i-1}^+(t)/g_i^-(t) = \exp \{ \beta[U_{i-1}(t)-U_i(t)]\}$
is satisfied exactly and the time-independence of $g_i^+(t) + g_i^-(t) = \nu_i^{\alpha} = \mathrm{const}$
is also maintained for small $\Delta x$ and a sufficiently smooth potential.
We have used here the notation $U_i(t) \equiv U(i \Delta x,t)$ and 
$U_{i \pm 1/2}(t) \equiv U(i \Delta x \pm \Delta x /2,t)$; 
$\beta = k_B T$ is the inverse of temperature and $\kappa_{\alpha}$ free fractional 
diffusion coefficient with dimension $\mathrm{cm}^2 \mathrm{s}^{-\alpha}$.
By passing to the continuous space limit $\Delta x\to 0$ 
as in Ref.~\cite{letter}, one finally obtains,
\begin{eqnarray} \label{FFPEmod}
\frac{ \partial }{\partial t} P (x, t) = 
\left [ - \frac{\partial}{\partial x} \frac{F (x, t)}{\eta _\alpha} 
+ \kappa _\alpha \frac{\partial ^2}{\partial x^2} \right ] 
\sideset{_0}{_t}{\mathop{\hat D}^{1 - \alpha}} P (x, t) \, .
\end{eqnarray}
In the latter equation $\eta _\alpha = ( \beta \kappa_{\alpha})^{-1}$ 
is the fractional friction coefficient.

In the following we use Eq.~(\ref{FFPEmod}) to study analytically 
the subdiffusion in time-periodic rectangular fields. 
Our study is complemented by stochastic simulations of the
underlying CTRW using the algorithm detailed in Ref.~\cite{heinsalu2006b}.

\begin{figure}[t]
\centering
\includegraphics[width=7.5cm]{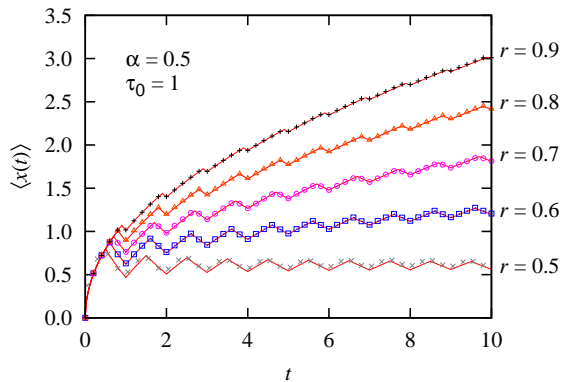}
\caption{(Color online) 
Average particle position $\langle x (t) \rangle$ for various values of 
the parameter $r$ (average force $\bar{F}$) and anomalous exponent $\alpha $.
Symbols represent the results from the numerical simulations of the CTRW obtained by 
averaging over $10 ^5$ trajectories (for $r = 0.5$ over $10 ^6$ trajectories). 
Continuous lines represent the analytical solution (\ref{xx}) of the FFPE (\ref{FFPEmod}). 
The time-period of the rectangular force (\ref{Fper}) is 
$\tau _0 = 1$ and the fractional exponent $\alpha = 0.5$;  
in numerical simulations $F_0 / (\eta _{\alpha} \sqrt{\kappa _{\alpha}}) = 1$ is used. 
The value $r = 0.9$ corresponds to $\bar{F} = 0.8 F_0$ and $\sigma = 0.6 F_0$;
$r = 0.8$ corresponds to $\bar{F} = 0.6 F_0$ and $\sigma = 0.8 F_0$; 
$r = 0.7$ corresponds to $\bar{F} = 0.4 F_0$ and $\sigma\approx 0.92 F_0$; 
$r = 0.6$ corresponds to $\bar{F} = 0.2 F_0$, $\sigma\approx 0.98 F_0$; 
and $r = 0.5$ to $\bar{F} = 0$, $\sigma = F_0$.} 
\label{Fig1}
\end{figure}
%


\section{Driven subdiffusion} \label{results}


We consider a dichotomous modulation  of a
biased subdiffusion where the absolute value of the bias is
fixed but its direction flips periodically in time, i.e.
\begin{equation} \label{Fper}
F (t) = F_0 \xi (t)
\end{equation}
with
\begin{eqnarray} \label{FperPM}
\xi (t) = \left\{ \begin{array}{ll} + 1 & \textrm{~for}~~ n \tau_0 < t < (n+r) \tau_0 \\
- 1 & \textrm{~for}~~ (n+r) \tau_0 < t < (n+1) \tau _0
\end{array} \right. .
\end{eqnarray}
Here $\tau _0$ is the period of the time-dependent force and $n = 0, 1, 2 \ldots $ 
The quantity $r \in (0,1)$ determines the value of the average force: 
\begin{equation}
\bar{F} = \langle F (t) \rangle _{\tau _0} = F_0 (2r - 1) \, .
\end{equation}
For $r = 0.5$ the average bias is zero and we recover the model investigated
in Ref.~\cite{heinsalu2007b}. 
Notice that the force $F(t)$ can be decomposed in the following way: $F(t) = \bar{F} + \tilde F(t)$. 
The asymmetric driving, 
\begin{eqnarray} \label{Fper2}
\tilde F (t) \!= \!
\left\{\!\begin{array}{ll} 2 F_0 (1-r) & \textrm{~for}~~ n \tau_0 < t < (n+r) \tau_0 \\
- 2 F_0 r & \textrm{~for}~~ (n+r) \tau_0 < t < (n+1) \tau _0
\end{array} \right. \!,
\end{eqnarray}
has a zero mean value, $\langle \tilde F (t) \rangle _{\tau _0} = 0$, 
and the driving root-mean-squared (rms) amplitude is 
$\sigma = \langle \tilde F^2 (t) \rangle^{1/2} _{\tau _0} = 2F_0 \sqrt{r(1-r)}$.
For a fixed average bias $\bar{F}$, this yields
\begin{eqnarray}
\sigma = 2 \bar{F} \frac{\sqrt{r(1-r)}}{2r-1}
\end{eqnarray}
and therefore one can vary the ratio $\sigma/ \bar{F}$ between $0$ for
$r=1$ and $\infty$ for $r=1/2+\epsilon$, $\epsilon\to 0$.
This offers the way to study the influence of an asymmetric, zero-mean
driving $\tilde{F}(t)$ with period $\tau_0$ and rms amplitude $\sigma$ 
on the subdiffusion under constant bias $\bar{F}$.

Let us begin by finding the recurrence relation  for the moments $\langle x ^n (t) \rangle$.  
Assuming in Eq.~(\ref{FFPEmod}) the force of the form 
(\ref{Fper}) with (\ref{FperPM}), multiplying both sides of Eq.~(\ref{FFPEmod}) by 
$x ^n$, and integrating over the $x$-coordinate one obtains,
\begin{eqnarray} \label{a2}
\frac{d \langle x ^n (t) \rangle}{d t} 
&=& n v _\alpha \xi (t) \sideset{_0}{_t}{\mathop{\hat D}^{1 - \alpha}} \langle x^{n-1} (t) \rangle \nonumber \\
&+& n (n - 1) \kappa _\alpha \sideset{_0}{_t}{\mathop{\hat D}^{1 - \alpha}} \langle x ^{n - 2} (t) \rangle \, ,
\end{eqnarray}
with subvelocity $v _{\alpha} = F_0 / \eta _{\alpha}$ ($n > 1$). 
For $n = 1$ the last term on the right hand side of Eq.~(\ref{a2}) is absent,
\begin{eqnarray} \label{xav}
\frac{d \langle x(t) \rangle}{d t} = \frac{v _\alpha}{\Gamma (\alpha)} 
\, \xi (t) \, t ^{\alpha - 1} \, .
\end{eqnarray}
Equations (\ref{a2}) and (\ref{xav}) will be used to calculate the 
average particle position and the mean square displacement.

\begin{figure}[t]
\centering
\includegraphics[width=7.5cm]{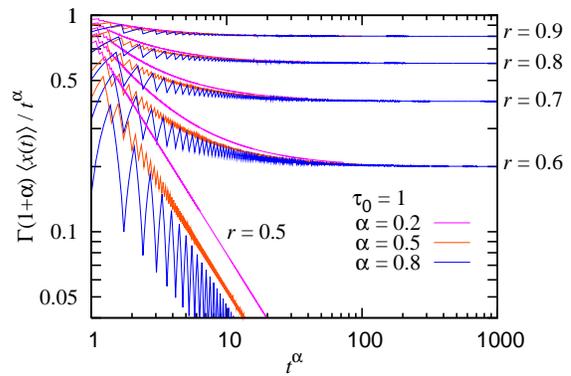}
\caption{(Color online) 
The analytical solution (\ref{xx}) for average particle position 
$\langle x (t) \rangle$ obtained from the FFPE (\ref{FFPEmod}) is presented 
for various values of the parameter $r$ (average bias $\bar{F}$) and anomalous exponent $\alpha $. 
The time-period of the force is $\tau _0 = 1$, however, in the long time limit 
the same asymptotic value is obtained for any value of $\tau _0$. 
The relation between $r$ and $\bar{F}$ and $\sigma$  is the same as in Fig.~\ref{Fig1}.} 
\label{Fig2}
\end{figure}
%


\subsection{Average particle position}


Upon integrating Eq.~(\ref{xav}) in time with $\xi (t)$ given by (\ref{FperPM}), 
the solution for the average particle position reads:
\begin{eqnarray} \label{xx}
&& \langle x (t) \rangle \nonumber \\
&& =\left\{ 
\begin{array}{l@{\quad \quad}}
x _N + \frac{v _{\alpha} t ^\alpha}{\Gamma (\alpha + 1)} \, , \quad N \tau _0 \le t < (N + r) \tau _0 \\
x _N' - \frac{v _{\alpha} t ^\alpha}{\Gamma (\alpha + 1)} \, , \quad (N + r) \tau _0 \le t < (N + 1) \tau _0 
\end{array} 
\right.
\end{eqnarray}
with
\begin{eqnarray} \label{xn1}
x _N &=& \langle x (0) \rangle - \frac{v _{\alpha} (N \tau _0) ^\alpha}{\Gamma (\alpha + 1)} 
+ \frac{v _{\alpha} \tau _0 ^\alpha}{\Gamma (\alpha + 1)} \nonumber \\
&\times& \sum_{n = 0} ^{N - 1} \left [2(n + r) ^\alpha -n ^\alpha - (n + 1) ^\alpha \right ] \, , \\
x _N' &=& x _N + \frac{2 v _{\alpha} \tau _0 ^\alpha}{\Gamma (\alpha + 1)} (N + r) ^\alpha \, ;
\end{eqnarray}
$N$ counts the number of time periods passed.

When the average bias is zero, i.e., $r = 0.5$, in the long time limit
the mean particle position approaches the constant value
\begin{eqnarray} \label{asympt}
\langle x (\infty) \rangle = v _{\alpha} \tau _0 ^{\alpha} b (\alpha ) / \Gamma(\alpha + 1),
\end{eqnarray}
with $b (\alpha ) = \sum _{n = 0} ^{\infty}[2 (n + 1/2) ^{\alpha} - n ^{\alpha} - (n + 1) ^{\alpha}]$. 
The function $b (\alpha )$ changes monotonously from $b (0) = 1$ to $b (1) = 0$. 
It describes the initial field phase effect which the system remembers forever when 
$\alpha \in (0, 1)$ (see also Ref.~\cite{heinsalu2007b}). 
This is one of the main differences between the anomalous motion in the absence 
of a force and in the presence of a time-dependent field with zero average value. 

In Fig.~\ref{Fig1} the analytical solution (\ref{xx}) for the mean
particle position $\langle x (t) \rangle$, obtained from the
FFPE (\ref{FFPEmod}), is compared with the numerical solution of the CTRW
for different values of $r$, i.e. for different values of the average bias $\bar{F}$.  
In Fig.~\ref{Fig2} the solution (\ref{xx}) is presented in the 
long time limit for various values of $r$ and $\alpha $. 
Figures \ref{Fig1} and \ref{Fig2} demonstrate that in the presence of an
average bias the mean particle position grows as $t ^\alpha$.
For all values of $r$, the asymptotic value of subvelocity corresponds
to the averaged bias $\bar{F} = F_0 (2r-1)$, indicating that the periodic
unbiased field does not affect the subdiffusion current for different values
$\bar{F}$ and the field rms-amplitude $\sigma$.

Furthermore, the results depicted in Figs.~\ref{Fig1} and \ref{Fig2} clearly show the 
phenomenon of the ``death of linear response'' of the fractional kinetics to time-dependent 
fields in the limit $t \to \infty$: the amplitude of the oscillations decays to
zero as $1 / t ^{1 - \alpha}$, Eq.~(\ref{xav}) (see also Refs.~\cite{Barbi, SokolovKlafter06}).  
The amplitude of the oscillations is larger for larger values of 
$\alpha $ and of $\tau _0$.
However, $\Gamma (1+\alpha) \langle x(t) \rangle / t^\alpha$ reaches asymptotically 
the same value for any $\alpha$ and $\tau _0$.


\subsection{Mean square displacement}


Let us now study the mean square displacement, defined as
\begin{equation} \label{MSD}
\langle \delta x ^2 (t) \rangle = \langle x ^2 (t) \rangle - \langle x (t) \rangle ^2 \,.
\end{equation}
For $n = 2$ one obtains from Eq.~(\ref{a2}),
\begin{equation} \label{AP1}
\frac{d \langle x ^2 (t) \rangle}{d t} 
= 2 v _\alpha \xi (t) \sideset{_0}{_t}{\mathop{\hat D}^{1 - \alpha}} \langle x (t) \rangle 
+ \frac{ 2 \kappa _\alpha }{\Gamma (\alpha)} \, t ^{\alpha - 1} \, .
\end{equation}
In order to find the analytical solution for the mean square displacement, 
we use the Laplace-transform method and the Fourier series expansion for 
$\xi (t) = \xi (t + \tau _0)$ given wuth Eq.~(\ref{FperPM}),
\begin{equation} \label{AP3}
\xi (t) = \sum_{n = - \infty}^{\infty} f _n \exp (i n \omega _0 t) \, ,
\end{equation}
with
\begin{eqnarray} \label{AP4}
f _n &=& \frac{1}{\tau _0} \int _{0} ^{\tau _0} \xi (t) \exp ( - in \omega _0 t) dt \nonumber \\
&=& [1 - \exp ( - inr 2 \pi )] / ( in \pi )
\end{eqnarray}
and $\omega _0 = 2 \pi / \tau _0$. 
Applying them to Eq.~(\ref{AP1}) and assuming $\langle x (0) \rangle = 0$ and 
$\langle x ^2 (0) \rangle = 0$ we obtain that in the long time limit 
(see Appendix~\ref{APPENDIX-1}),
\begin{eqnarray} \label{MSD-1}
\langle x ^2 (t) \rangle &=& \frac{2v _\alpha ^2 (2r - 1) ^2 }{\Gamma (2 \alpha + 1)} t ^{2 \alpha } 
+ \frac{2 \kappa _\alpha}{\Gamma (\alpha + 1)} t ^\alpha \nonumber \\
&+& \frac{2 v _\alpha ^2 (2r - 1) S_1}{\omega _0 ^\alpha \Gamma (\alpha + 1)} t ^\alpha
 + \frac{8v _\alpha ^2 \cos (\alpha \pi / 2)}{\pi ^2 \omega _0 ^\alpha \Gamma (\alpha + 1)}  \nonumber \\
&\times & \left [\zeta (2 + \alpha ) - \sum _{n = 1} ^{\infty } \frac{\cos (nr 2 \pi )}{n ^{2 + \alpha }} \right ] t ^\alpha \, ;
\end{eqnarray}
here $\zeta (x)$ is the Riemann's zeta-function and $S_1$ is a function of 
$\alpha$ and $r$ as given by Eq.~(\ref{AP13}) in Appendix~\ref{APPENDIX-1}.

\begin{figure}[t]
\centering
\includegraphics[width=7.5cm]{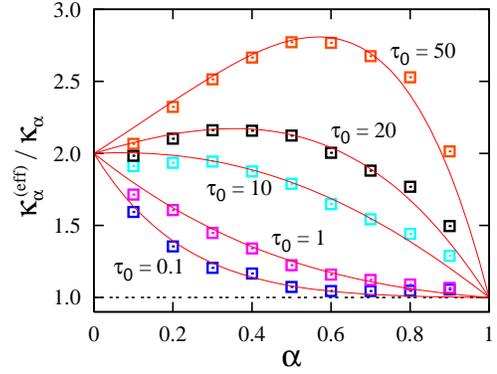}
\caption{(Color online) Scaled effective fractional diffusion coefficient $\kappa _\alpha ^{\mathrm{(eff)}}$ 
versus fractional exponent $\alpha$ for different driving periods $\tau _0$. 
The average bias is zero ($r = 0.5$). 
The analytical prediction (\ref{basic}) (continuous lines) is compared with the results (symbols) 
obtained from the numerical simulation of the CTRW by averaging over $10 ^5$ trajectories. 
For $\tau _0 > \tau _0^* \approx 8.818$ \cite{heinsalu2007b} the effective fractional diffusion coefficient 
$\kappa _\alpha ^{\mathrm{(eff)}}(\alpha )$ exhibits a maximum.}
\label{Fig3}
\end{figure}

For $r = 0.5$ (average zero bias) the first and third term in Eq.~(\ref{MSD-1}) are equal to zero. 
Furthermore, in the long time limit the average particle position $\langle x (\infty ) \rangle $ is a finite constant. 
The asymptotic behavior of the mean square displacement is thus proportional to $t ^{\alpha}$ as in the force
free case, however, characterized by an effective fractional diffusion coefficient 
$\kappa _{\alpha} ^\mathrm{(eff)}$ instead of the free fractional diffusion coefficient $\kappa _\alpha$, i.e.
$\langle \delta x ^2 (t) \rangle = 2\kappa _{\alpha} ^\mathrm{(eff)} t^ {\alpha} / \Gamma (1 + \alpha )$ for $t \to \infty$. 
The effective diffusion coefficient is \cite{heinsalu2007b},
\begin{eqnarray} \label{basic}
\kappa _\alpha ^{\mathrm{(eff)}} &=& \kappa _\alpha 
+ \frac{8 F_0^2}{\pi ^2 \eta _\alpha ^2 \omega _0 ^\alpha } \, \zeta (\alpha + 2) \nonumber \\
&\times& \left(1- \frac{1}{2 ^{\alpha + 2}} \right ) \cos (\alpha \pi / 2) \, .
\end{eqnarray}
The driving-induced part of the effective subdiffusion coefficient is directly proportional 
to the square of driving amplitude $F_0$ and inversely proportional to $\omega _0 ^\alpha $. 
For slowly oscillating force fields this leads to a profound 
acceleration of subdiffusion compared with the force free case: 
an optimal value of the fractional exponent $\alpha$ exists, at which the driving-induced part 
of the effective fractional diffusion coefficient possesses a maximum (see Fig.~\ref{Fig3}).

When $r \neq 0.5$ (finite average force) we obtain in the long time limit
for $\langle x (t) \rangle ^2$ (see Appendix~\ref{APPENDIX-2}),
\begin{equation} \label{MSD-2}
\langle x (t) \rangle ^2 = \frac{v _\alpha ^2 (2r - 1) ^2}{\Gamma ^2 (\alpha + 1)} t ^{2 \alpha } + 
\frac{2v _\alpha ^2 (2r - 1) S_1}{\omega _0 ^\alpha \Gamma (\alpha + 1)} t ^\alpha
\end{equation}
[$S_1$ is given by Eq.~(\ref{AP13}) in Appendix~\ref{APPENDIX-1}].
Clearly, the leading term in Eq.~(\ref{MSD-2}) corresponds to subvelocity in constant 
field $\bar{F}$ (averaged bias), i.e. the influence of periodic, unbiased driving 
$\tilde F(t)$ dies asymptotically out, as illustrated in Fig.~\ref{Fig2}.

The results (\ref{MSD-1}) and (\ref{MSD-2}) indicate that in the 
presence of a rectangular time-periodic force with a finite average value the
general behavior of the mean square displacement is similar to the
case of a constant force, i.e. the mean square displacement 
$\langle \delta x ^2 (t)\rangle $ consists of terms proportional to 
$t ^\alpha $ and $t ^{2 \alpha }$.
In fact, for the leading term proportional to $t ^{2 \alpha }$ in the 
mean square displacement one obtains the coefficient
\begin{equation*} \label{AP22}
\frac{{\bar{F}}^2}{\eta _\alpha ^2} \left [ \frac{2}{\Gamma (2 \alpha + 1)} 
- \frac{1}{\Gamma ^2 (\alpha + 1)} \right ] \, .
\end{equation*}
This coefficient is the same as in the case of the subdiffusive motion under the influence 
of a constant force if the value of the constant force would be $\bar{F}$.

\begin{figure}[t]
\centering
\includegraphics[width=7.5cm]{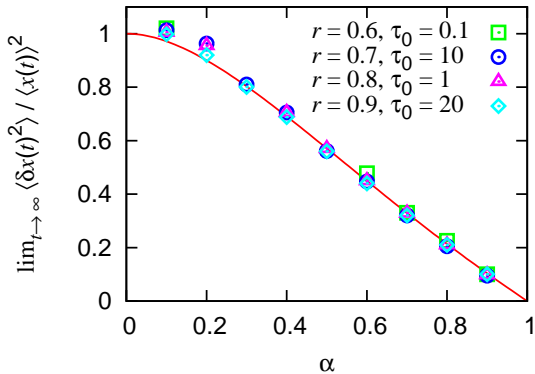}
\caption{(Color online) 
The asymptotic scaling relation (\ref{anomal}). 
Symbols correspond to the numerical results obtained from the CTRW 
for different values of $r$ ($r \neq 0.5$) and $\tau_0$.
Solid curve corresponds to the analytical result (\ref{anomal}). 
Different values of the parameter $r$ correspond to different
values of the bias $\bar{F}$ and the periodic field rms $\sigma$, as described in
Fig.~\ref{Fig1}. 
The field parameters do not influence the results within the statistical errors.} 
\label{Fig4}
\end{figure}

Furthermore, similarly to the case of a constant bias, the asymptotic scaling relation holds 
($r \neq 0.5 $, $\bar{F} \neq 0 $) between the mean square displacement and
average particle position.
In the limit $t \to \infty$ the mean square displacement grows as 
$\langle \delta x ^2 (t) \rangle \propto t ^{2 \alpha }$  and
\begin{eqnarray} \label{anomal}
\lim_{t \to \infty} \frac{\langle \delta x^2 (t) \rangle}{ \langle x (t) \rangle ^2} 
= \frac{2 \Gamma ^2 (\alpha + 1)}{\Gamma (2 \alpha + 1)} - 1 \, , \quad \bar{F} \neq 0 \, .
\end{eqnarray}
It is illustrated by Fig.~\ref{Fig4},  where the analytical curve
[Eq.~(\ref{anomal})] is compared with the numerical results. 
The universality of relation (\ref{anomal}) under the unbiased driving $\tilde F(t)$
means that the biased diffusion is not affected by the driving. 
This is in a sharp contrast with the unbiased diffusion in Fig.~\ref{Fig3}.


\section{Conclusion}


With this work we presented the derivation of the FFPE (\ref{FFPEmod}) for a 
special class of space- and time-dependent force fields from the underlying CTRW picture. 
Our derivation shows along with the corresponding discussion
that it is difficult to justify this equation for time-dependent forces 
different from $F(x, t) = F(x) \xi (t)$ with $\xi (t) = \pm 1$
beyond the linear response approximation. 
Using the FFPE (\ref{FFPEmod}) we demonstrated that the universal 
scaling relation (\ref{anomal}) for a biased subdiffusion is not affected 
by the additional action of a time-periodic zero-mean rectangular driving;
neither is the asymptotic anomalous current nor the anomalous biased diffusion.
We argue that this result is general and it is valid for other driving forms
as well.
This driving-immunity is due to the fact that the CTRW subdiffusion occurs 
in a random operational time which lacks mean value, whereas any physical field
changes in the real, physical time. 
The CTRW-based subdiffusion fails to respond 
asymptotically to such time-dependent fields while on its intrinsic random 
operational time scale any real, alternating field is acting infinitely fast
and it makes effectively no influence in a long run \cite{heinsalu2007b,goychuk07},
unless the rate of its change is precisely zero. 
This is the main reason for the observed anomalies. 
The remarkable enhancement of the unbiased subdiffusion within the CTRW framework 
by time-periodic rectangular fields is rather an exception than the rule.


\begin{acknowledgements}


This work has been supported by the targeted financing project SF0690030s09, 
Estonian Science Foundation via grant no. 7466 (M.P., E.H.), 
Spanish MICINN and FEDER through project FISICOS (FIS2007-60327) (E.H.), 
the EU NoE BioSim, LSHB-CT-2004-005137 (M.P.),
the DFG-SFB-486, and by the Volks\-wagen Foundation, no. I/80424 (P.H.).
\end{acknowledgements}


\appendix


\section{} \label{APPENDIX-1}


Using the property 
$\mathcal{L} \{ d f(t) /dt \} = s \mathcal{L} \{ f(t) \} - f(0)$ 
and assuming the initial conditions
$\langle x(t) \rangle = 0$ and $\langle x^2 (t) \rangle = 0$ 
we obtain from Eq.~(\ref{AP1}),
\begin{equation} \label{AP2}
\mathcal{L}\{ \langle x^2 (t) \rangle \} 
= \frac{2 \kappa_\alpha}{s^{\alpha +1}} 
+ \frac{2v_\alpha}{s}\mathcal{L} \{ \xi (t) \sideset{_0}{_t}{\mathop{\hat D}^{1-\alpha}} \langle x(t) \rangle \} \, .
\end{equation}
Considering that
\begin{equation*}
\mathcal{L}\{ \sideset{_0}{_t}{\mathop{\hat D}^{1-\alpha}} \langle x(t) \rangle \} 
= s^{1-\alpha} \mathcal{L}\{ \langle x(t) \rangle \} 
= s^{-\alpha} \mathcal{L}\{d \langle x(t) \rangle /dt \} \, ,
\end{equation*}
one obtains,
\begin{eqnarray} \label{AP5}
\mathcal{L} \{ \xi(t) \sideset{_0}{_t}{\mathop{\hat D}^{1-\alpha}}
\langle x(t) \rangle \} &=& \sum_{n=-\infty}^{\infty} \frac{f_n}{(s-in\omega_0)^\alpha } \nonumber \\
&\times & \mathcal{L}_{s-in\omega_0}\left \{\frac{ d \langle x(t) \rangle }{dt} \right \} \, .
\end{eqnarray}
Using Eq.~(\ref{xav}) with (\ref{AP3}), it follows,
\begin{equation} \label{AP6}
\mathcal{L}_{s-in\omega_0}\left \{\frac{ d \langle x(t) \rangle }{dt} \right \} 
= v_\alpha \sum_{m=-\infty}^{\infty} \frac{f_m}{[s-i\omega_0(n+m)]^\alpha} \, .
\end{equation}
Inserting (\ref{AP5}) and (\ref{AP6}) into Eq.~(\ref{AP2}) we obtain,
\begin{eqnarray} \label{AP7}
\mathcal{L}\{ \langle x^2(t) \rangle \} &=& \frac{2 \kappa _\alpha}{s^{\alpha +1}} 
+ \frac{2v_\alpha^2}{s} \sum_{n=-\infty}^{\infty} \frac{f_n}{(s-in\omega_0)^\alpha} \nonumber \\
&\times & \sum_{m=-\infty}^{\infty} \frac{f_m}{[s-i\omega_0(n+m)]^\alpha} \, .
\end{eqnarray}

Let us separate in Eq.~(\ref{AP7}) the terms $m=0$ and $n=0$,
\begin{eqnarray} \label{AP11}
&\mathcal{L}& \{ \langle x^2(t) \rangle \} =
\frac{2\kappa_\alpha}{s^{\alpha +1}} + \frac{2v_\alpha^2 f_0^2 }{s^{2\alpha +1}} \nonumber \\
&+& \frac{2v_\alpha^2 f_0 }{s^{\alpha +1}} \sum_{\substack{m=-\infty
\\ m\neq 0}}^{\infty} \frac{f_m}{(s-im\omega_0)^\alpha } +
\frac{2v_\alpha ^2}{s} \sum_{\substack{n=-\infty \\
n\neq
0}}^{\infty} \frac{f_n}{(s-in\omega_0)^\alpha } \nonumber \\
&\times & \sum_{m=-\infty}^{\infty}
\frac{f_m}{[s-i\omega_0(n+m)]^\alpha} \, .
\end{eqnarray}
In the long time limit, i.e., in the limit $s \to 0$, in the
double sum only terms with $m=-n$ contribute, giving thus,
\begin{eqnarray} \label{AP12}
\mathcal{L} \{ \langle x^2(t) \rangle \} &=&
\frac{2\kappa_\alpha}{s^{\alpha +1}} + \frac{2v_\alpha^2 f_0^2
}{s^{2\alpha +1}} + \frac{2v_\alpha^2 f_0 }{\omega _0^\alpha
s^{\alpha +1}} \sum_{\substack{m=-\infty
\\ m\neq 0}}^{\infty} \frac{f_m}{(-im)^\alpha } \nonumber \\
&+& \frac{2v_\alpha^2}{\omega _0^\alpha s^{\alpha +1}} 
\sum_{\substack{n=-\infty \\ n\neq 0}}^{\infty} \frac{|f_n|^2}{(-in)^\alpha } \, .
\end{eqnarray}
Let us compute the sums. Considering that
\begin{equation*}
\sum_{\substack{m=-\infty
\\ m\neq 0}}^{\infty} \frac{f_m}{(-im)^\alpha } 
= \sum _{m=1}^{\infty}\left [\frac{f_m}{(-im)^\alpha} + \frac{f_{-m}}{(im)^\alpha } \right ] \, ,
\end{equation*}
and replacing here $f_m $ from (\ref{AP4}), one obtains,
\begin{eqnarray} \label{AP13}
\sum_{\substack{m=-\infty \\ m\neq 0}}^{\infty}
\frac{f_m}{(-im)^\alpha } = \left. \frac{2}{\pi } \right[ \zeta (1+\alpha ) \sin{(\alpha \pi /2)} \nonumber \\
- \sum_{m=1}^{\infty } \left. \frac{\sin{[(\alpha -4mr)\pi / 2]}}{m^{1+\alpha }}  \right] \equiv S_1 \, ;
\end{eqnarray}
here $\zeta (x)$ is the Riemann's zeta-function.
Analogously,
\begin{eqnarray} \label{AP14}
\!\!\!\! \sum_{\substack{n=-\infty \\ n\neq 0}}^{\infty} \! \frac{|f_n|^2}{(-in)^\alpha } \!\! 
&=& \!\! \frac{4}{\pi ^2 } \cos(\alpha \pi / 2) \nonumber \\
&\times & \!\!\! \left[ \zeta (2+\alpha ) - \sum_{n=1}^{\infty } \frac{\cos(nr2\pi )}{n^{2+\alpha }}  \right] \, .
\end{eqnarray}
Replacing these sums into Eq.~(\ref{AP12}) and considering that $f_0 = 2r-1$ 
[see (\ref{AP4}) together with (\ref{FperPM})], we get,
\begin{eqnarray} \label{AP15}
\mathcal{L} \{ \langle x^2(t) \rangle \} 
= \frac{2v_\alpha^2 (2r-1)^2 }{s^{2\alpha +1}} + \frac{2\kappa_\alpha}{s^{\alpha +1}}
+  \frac{2 v_\alpha^2 (2r-1) S_1}{\omega _0^\alpha s^{\alpha +1}} \nonumber \\
+ \frac{8v_\alpha^2}{\pi ^2 \omega _0^\alpha s^{\alpha +1}} \cos(\alpha \pi / 2) 
\left[ \zeta (2+\alpha ) - \sum_{n=1}^{\infty } \frac{\cos(nr2\pi )}{n^{2+\alpha }}  \right] \, . \nonumber \\
\end{eqnarray}
Taking here the inverse Laplace transform one obtains the expression for 
$\langle x^2(t)\rangle $ in the long time limit [Eq.(\ref{MSD-1})].


\section{} \label{APPENDIX-2}


Using Eq.~(\ref{xav}), the quantity $\langle x(t) \rangle ^2 $ can be written in the following way,
\begin{equation} \label{AP8}
\langle x(t) \rangle ^2 
= \frac{2v_\alpha^2}{\Gamma ^2(\alpha)} \int _0^t dt' \int_0^{t'} \xi (t') t'^{\alpha -1} \xi(t'') t''^{\alpha -1} dt'' \, .
\end{equation}
Exploiting the property $\mathcal{L}\{ \int_0^t f(t') dt' \} = s^{-1} \mathcal{L} \{ f(t') \}$ 
and denoting $t'=t$ and $t''=t'$, we can write,
\begin{eqnarray} \label{AP9}
&\mathcal{L}& \{ \langle x(t) \rangle ^2 \} =
\frac{2v_\alpha^2}{\Gamma ^2(\alpha) s} \mathcal{L} \left \{ \xi (t) t^{\alpha -1} \int_0^t \xi(t') t'^{\alpha -1} dt' \right \}
\nonumber \\
&=& \frac{2v_\alpha^2}{\Gamma ^2(\alpha) s} \sum_{n=-\infty}^{\infty} f_n \sum_{m=-\infty}^{\infty} f_m \nonumber \\
&\times & \mathcal{L}_{s-in\omega_0} \left \{ t^{\alpha -1} \int _0^t t'^{\alpha -1} \exp(im\omega_0t') dt' \right \} \, .
\end{eqnarray}

For $\alpha =1$ the latter equation gives,
\begin{eqnarray} \label{AP10}
\mathcal{L} \{ \langle x(t) \rangle ^2 \} &=&
\frac{2v_\alpha^2}{s} \sum_{n=-\infty}^{\infty} \frac{f_n}{s-in\omega_0} \nonumber \\
&\times & \sum_{m=-\infty}^{\infty} \frac{f_m}{s-i\omega_0(n+m)} \, .
\end{eqnarray}
Comparing this result with Eq.~(\ref{AP7}) we see that
$\mathcal{L} \{ \langle \delta x^2 (t)\rangle \} = 2 \kappa_\alpha/s^2$, and thus 
$\langle \delta x^2 (t)\rangle = 2 \kappa_\alpha t$, as it should be for normal Brownian motion.

For $0 < \alpha < 1$ it is more convenient to proceed as follows.
Let us calculate the Laplace trasform of $\langle x(t) \rangle$.
Considering that $\mathcal{L} \{ \langle x(t) \rangle \} = s^{-1} \mathcal{L} \{ d \langle x(t) \rangle / dt \}$ one obtains [see (\ref{AP6})],
\begin{eqnarray} \label{}
\mathcal{L} \{ \langle x(t) \rangle \} &=& v_\alpha \sum _{n=-\infty}^{\infty} \frac{f_n}{s(s-in\omega _0)^\alpha} \nonumber \\
&=& \frac{v_\alpha f_0}{s^{1+\alpha}} + v_\alpha \sum _{\substack{n=-\infty \\ n\neq 0}}^{\infty} \frac{f_n}{s(s-in\omega _0)^{\alpha}} \, .
\end{eqnarray}
In the limit $t \to \infty$, i.e. $s \to 0$, the latter equation becomes,
\begin{equation} \label{}
\mathcal{L} \{ \langle x(t) \rangle \} = \frac{v_\alpha f_0}{s^{1+\alpha}} 
+ \frac{v_\alpha}{s} \sum _{\substack{n=-\infty \\ n\neq 0}}^{\infty} \frac{f_n}{(-in\omega _0)^{\alpha}} \, .
\end{equation}
Taking into account (\ref{AP13}) we can write,
\begin{equation} \label{}
\mathcal{L} \{ \langle x(t) \rangle \} = \frac{v_\alpha f_0}{s^{1+\alpha}} 
+ \frac{v_\alpha S_1}{\omega _0 ^\alpha s}  \, .
\end{equation}
Taking here the inverse Laplace transform we have for $t \to \infty$,
\begin{equation} \label{}
\langle x(t) \rangle = \frac{v_\alpha (2r -1)}{\Gamma (\alpha +1)} \, t^\alpha 
+ \frac{v_\alpha S_1}{\omega _0 ^\alpha}  \, .
\end{equation}
From here one obtains the result (\ref{MSD-2}) for $\langle x(t) \rangle ^2$.



\end{document}